\documentclass[12pt,a4paper]{article}
\usepackage{amsmath}
\usepackage{amsfonts}
\usepackage{amssymb}
\usepackage{graphicx}

\begin{document}

\author{Vladimir K. Petrov}
\title{On the possibility of the critical behavior of LGT in the area of
asymptotically large $\beta$.}
\date{}
\maketitle
\begin{abstract}
Coupling dependence on lattice spacing and size is estimated analytically at
$\beta\rightarrow\infty$ region where for $a\rightarrow0$ the critical area is
shifted in accordance with Callan-Symanzik relation. In considered
approximation no trace of critical behavior is found in this area.
\end{abstract}

\section{Introduction}

In lattice gauge theory (LGT) the non-perturbative aspects are of primary
interest, but the renormalization-group technique is added, as a rule, in a
perturbative way and the relation between cut-off $\Lambda_{L}$ and gauge
coupling is given by (see e.g. \cite{ekr,b-p95})
\begin{equation}
a\Lambda_{L}~\equiv R(\beta)=\exp\left\{  -\dfrac{\beta}{4Nb_{0}}+\frac{b_{1}%
}{2b_{0}^{2}}\ln\frac{\beta}{2Nb_{0}}\right\}  ,\label{bet}%
\end{equation}
where $\beta=2Ng^{-2}$, $g$ is coupling constant, $a$ is lattice spacing, $N $
is the number of colors and
\begin{equation}
b_{0}=\frac{11}{3}\frac{N}{16\pi^{2}};~~~b_{1}=\frac{34}{3}\left(  \frac
{N}{16\pi^{2}}\right)  ^{2}.\label{ren}%
\end{equation}

Although the continuum limit in asymptotically free theories corresponds to
$g\rightarrow0,$ there are reasons to believe, that such theories do not
become perturbative at $a\sim0$ \cite{hasenfratz}. On the basis of today's
numeric computations, it is difficult to anticipate the behavior of $R(\beta)$
in the limit of$\ a\rightarrow0$, taking perturbative calculations as a
guidance. Indeed, numerical studies \cite{b-p95} showed deviations from
$\left(  \ref{bet}\right)  $ when the correlation length begins to grow. It is
especially worth to note, that these deviations are of such a pattern, as if
the theory approaches the fixed point $g_{0}$ at which the Callan-Symanzik
$\beta$-function $~\beta_{CS}(g_{0})\ $has $n$-order zero
\begin{equation}
\beta_{CS}(g)\equiv-a\frac{\partial g}{\partial a}\simeq-b_{n}^{\prime}%
(g^{2}-g_{0}^{2})^{n}\label{fp}%
\end{equation}
and consequently the theory is not asymptotically free. Cogent arguments in
favor of such behavior of $~\beta_{CS}(g_{0})\ $were given in \cite{PS}.

Data on deep inelastic scattering does not eliminate the fixed point
\cite{ss95}, however, the available data cannot distinguish between the first
and second order fixed points. One may conclude only, that $\beta_{CS}(g)\ $
may, indeed, turn into zero, presumably located within intervals
$0.8<g_{0}<1.\,\allowbreak2$ for $n=1$ or $0.6<g_{0}<1$ for $n=2$. Such
intervals may appear even wider, but beyond specified intervals the errors
doesn't allow to determine$~g_{0}$ from the data \cite{ss95}.

Phenomenological analysis of available MC lattice data in the $SU(2)$%
-gluodynamics shows no contradiction with the \textit{first} order fixed point
of $\beta_{CS}(g)$ \cite{bgk}. In case of $SU(2)$-gluodynamics the presumed
fixed point may located at $g_{0}~\simeq0.563$ \cite{bgk}.

On the other hand, it is quite within a reason to suggest that the lattice
data deviations from $\left(  \ref{bet}\right)  $ are the result of the finite
size effects. The dependence on spatial lattice size $N_{\sigma}$ may be
almost removed, e.g., for the $SU(3)$ gauge theory \cite{b-p95}, by
\begin{equation}
\beta_{c}(N_{\tau},N_{\sigma})=\beta_{c}(N_{\tau},\infty)-\left(  N_{\tau}%
{/}N_{\sigma}\right)  ^{3}h\label{kc}%
\end{equation}
with $h\lesssim0$.$1$ \cite{Iwa92}. To remove remnant deviation, one may
assume, that $\left(  \ref{bet}\right)  $ contain some preasymptotic terms,
which cause a disagreement with the data at achieved $\beta$ and vanish when
$\beta\rightarrow\infty$. Indeed, a set of sophisticated tunings, such as
$R(\beta)\rightarrow\lambda(\beta)R(\beta)$ and $\Lambda_{L}$ $\rightarrow
\Lambda_{L}(\beta)$ with $\lambda(\infty)=1$ and $\Lambda_{L}(\infty)=const$
\cite{ekr,b-p95} may bring $\left(  \ref{bet}\right)  $ into sync with
available MC data.

We see that both approaches have enough room for adjustment and their capacity
in experimental data description will be hardly exhausted in foreseeable
future. We would like to try analytical estimations to find more or less
tangible difference between such approaches. To study the continuum limit it
is more convenient to regard spacing $a$ as an independent parameter and
$\beta=\beta\left(  a\right)  $. As it follows from $\left(  \ref{fp}\right)
$ fixed pole model predicts $\lim_{a\rightarrow0}g\left(  a\right)  =g_{0}%
\sim1$. Unfortunately, in the area $g\sim1$ the analytical methods are not
efficient enough, but the perturbative approach $\left(  \ref{bet}\right)  $
dictates $\beta\propto-\ln a\Lambda_{L}$ in a limit $a\rightarrow0$. Moreover,
critical coupling $\beta_{c}$ corresponds to the critical temperature
$T_{c}=1/\left(  N_{\tau}a_{c}\right)  $, where $a_{c}$ is defined by
$\beta_{c}=\beta\left(  a_{c}\right)  $ and determined from $\left(
\ref{bet}\right)  $. Therefore, if one claims $T_{c}=T_{c}\left(  N_{\tau
}\right)  \rightarrow const$ and $\Lambda_{L}\rightarrow const$ with rising
temporal lattice size $N_{\tau}$, then for the critical coupling we get
\begin{equation}
\beta_{c}\simeq4Nb_{0}\ln\left(  N_{\tau}T_{c}/\Lambda_{L}\right)
\simeq4Nb_{0}\ln N_{\tau}.
\end{equation}
So the critical area is steadily displaced into the region of infinite $\beta$
in which one may expect essential simplification of analytical computations.
In this paper we try to at least partly clarify some features of critical
behavior in the area of $N_{\tau}$ $\rightarrow\infty$ and $\beta
\rightarrow\infty$ inaccessible in MC experiment. That allows us to establish
a relation between $\beta$ and $a$ and compare it with $\left(  \ref{bet}%
\right)  $.

\section{ Partition function at extremely large $\beta$}

Conjecturable vanishing $N_{\tau}$-dependance of $\beta_{c}$ in a fixed point
model \cite{bgk}
\begin{equation}
g_{c}~=g_{0}~+(3.15N_{\tau})^{-b_{1}^{\prime}}~;\qquad b_{1}^{\prime
}=0.111\label{fs}%
\end{equation}
looks reasonable for $Z(N)$ LGT where critical coupling $\beta_{c}(N_{\tau
},\infty)$ for $N=2,3$ tends to finite value $\beta_{c}(\infty,\infty)=\left(
1-1/N\right)  \ln\left(  1+\sqrt{N}\right)  $. However, such finite size
dependance is considered as unallowable for the standard $SU(N)$ LGT as
inconsistent with $\left(  \ref{bet}\right)  $. Moreover, vanishing of
$N_{\tau}$-dependance of the crossover peak may be regarded in $SU(N)$ LGT as
a reason to conclude that such crossover is not the result of an ordinary
phase transition \cite{es96}.

Such difference between $SU(N)$\ and $Z(N)$\ may be regarded as still another
reason to presume that the center subgroup loses its significance at
$\beta\rightarrow\infty$.\ It is generally assumed \cite{kt98} that, in the
continuum formulation there is no local distinction between pure $SU(N)$\ and
$SU(N)/Z(N)$, so it is expected that such difference will disappear in\ LGT as
the continuum is approached. Although some speculative reasons indicate that
$Z(N)$\ doesn't play any essential role when $a\rightarrow0$\ (and
consequently $\beta\rightarrow\infty$), yet it is not incurious to estimate
how fast the center contribution may fade out with increasing $\beta$.

Let us consider partition function of the pure Yang-Mills theory
\begin{equation}
Z=\int\exp\left\{  -S\right\}  \prod_{x;\nu}d\mu\left(  U_{\nu}\left(
x\right)  \right)  ;\quad U_{\nu}\left(  x\right)  \in SU\left(  N\right)  .
\end{equation}

As a rule, Wilson action contains the link variables $U_{\mu}\left(  x\right)
$ in fundamental representation
\begin{equation}
S=\sum_{x}\sum_{\mu>\nu}\mathcal{S}_{\mu\nu}\left(  x\right)
;\ \ \ \mathcal{S}_{\mu\nu}\left(  x\right)  =-\beta\tfrac{1}{N}%
\operatorname{Re}\chi\left(  U_{\mu\nu}\left(  x\right)  \right)  ;\label{S}%
\end{equation}
with$\ \chi\left(  U\right)  \equiv\mathrm{Tr}U$ and the plaquette variable
$U_{\mu\nu}\left(  x\right)  $ is defined as
\begin{equation}
U_{\mu\nu}\left(  x\right)  =U_{\mu}\left(  x\right)  U_{\nu}\left(
x+\mu\right)  U_{\mu}^{\dagger}\left(  x+\nu\right)  U_{\nu}^{\dagger}\left(
x\right)  ;\quad\mu,\nu=0,1,2,3.\label{p}%
\end{equation}

Following \cite{ukawa} we decompose $SU\left(  N\right)  $ matrices $U_{\mu
\nu}$ as
\begin{equation}
U_{\mu\nu}=z_{\mu\nu}\tilde{U}_{\mu\nu};
\end{equation}
where
\begin{equation}
z_{\mu\nu}\left(  x\right)  =z_{\mu}\left(  x\right)  z_{\nu}\left(
x+\mu\right)  z_{\mu}^{\ast}\left(  x+\nu\right)  z_{\nu}^{\ast}\left(
x\right)  \in Z\left(  N\right) \label{z}%
\end{equation}
and
\begin{equation}
-\tfrac{\pi}{N}<\arg\chi\left\{  \tilde{U}\right\}  <\tfrac{\pi}{N},\label{r}%
\end{equation}

Now \textit{plaquette} action $\mathcal{S}_{\mu\nu}\left(  x\right)  $ in
$\left(  \ref{S}\right)  $ may be rewritten as
\begin{equation}
-\mathcal{S}_{\mu\nu}\left(  x\right)  =\hat{\beta}_{\mu\nu}\left(  \tilde
{U}\left(  x\right)  \right)  z_{\mu\nu}\left(  x\right) \label{sz}%
\end{equation}
so we see that $\left(  \ref{sz}\right)  $ presents the action of $Z\left(
N\right)  $\ gluodynamics with varying couplings
\begin{equation}
\hat{\beta}_{\mu\nu}\left(  \tilde{U}\left(  x\right)  \right)  =\beta
\tfrac{1}{N}\operatorname{Re}\chi\left\{  \tilde{U}_{\mu\nu}\left(  x\right)
\right\} \label{coup}%
\end{equation}
In particular for $N=2$ gluodynamics in (3+1)-dimensional space we get
\begin{equation}
Z=\int d\tilde{U}\exp\left\{  -\sum_{x;\mu\nu}\widetilde{\mathcal{S}}_{\mu\nu
}\left(  x\right)  -\Xi\right\} \label{parti}%
\end{equation}
with
\begin{equation}
-\widetilde{\mathcal{S}}_{\mu\nu}\left(  x\right)  =\ln\left(  2\cosh
\hat{\beta}_{\mu\nu}\right)  -\beta
\end{equation}
and
\begin{equation}
\exp\left\{  -\Xi\right\}  =\sum_{\left(  z\right)  }\prod_{x;\mu\nu}\left(
1+z_{\mu\nu}\left(  x\right)  \tanh\hat{\beta}_{\mu\nu}\right)  .
\end{equation}

Duality transformation $z_{\mu\nu}\left(  x\right)  \rightarrow z_{\rho\omega
}^{\prime}\left(  x^{\prime}\right)  $ may be fulfilled for $Z\left(
N\right)  $\ gluodynamics with coupling being different for each plaquette
\cite{ukawa} and one may get
\begin{equation}
\exp\left\{  -\Xi\right\}  =\sum_{\left(  z^{\prime}\right)  }\left(
1+z_{\rho\omega}^{\prime}\left(  x^{\prime}\right)  \tanh\hat{\beta}%
_{\rho\omega}^{\prime}\right)  .\label{Z}%
\end{equation}
So the original plaquettes $z_{\mu\nu}\left(  x\right)  $ carrying coupling
$\hat{\beta}_{\mu\nu}\left(  \tilde{U}\left(  x\right)  \right)  $ are
transformed into dual $Z(2)$ plaquettes $z_{\rho\omega}^{\prime}\left(
x^{\prime}\right)  $ with varying coupling $\tilde{\beta}_{\rho\omega}%
^{\prime},$ that are related to $\hat{\beta}_{\mu\nu}\left(  \tilde{U}\left(
x\right)  \right)  $ by
\begin{equation}
\tanh\hat{\beta}_{\rho\omega}^{\prime}=\exp\left\{  -2\hat{\beta}_{\nu\mu
}\left(  \tilde{U}\left(  x\right)  \right)  \right\}  ,\quad\mu\neq\nu
\neq\rho\neq\omega.\label{dc}%
\end{equation}
Summing over the dual $Z(2)$ variables $\left\{  z^{\prime}\right\}  $ we
obtain
\begin{equation}
-\Xi=\sum_{A}\exp\left\{  -2\sum_{\left(  x\nu\mu\right)  ^{\prime}\in A}%
\hat{\beta}_{\nu\mu}\left(  \tilde{U}\left(  x\right)  \right)  \right\}
\label{kc-F}%
\end{equation}
where $\sum_{A}$ is taken over all closed self-avoiding\footnote{Although in
$SU(3)$ case surfaces $A$ are not obligatory self-avoiding, this doesn't make
essential change.} connected surfaces $A\ $and equiform surfaces considered as
different, if they are located at different places.

In the area $\beta\sim2N$ the critical behavior of the partition function is
defined mainly by $\Xi$ and, therefore, $SU\left(  N\right)  \simeq Z\left(
N\right)  $ approximation, i.e.
\begin{equation}
S\simeq\Xi;\qquad\hat{\beta}_{\nu\mu}\left(  \tilde{U}\left(  x\right)
\right)  \simeq\beta,\label{zN}%
\end{equation}
gives reasonable description of phase structure. Indeed, in the area
$g^{2}\sim1$ the center elements carry most of the information about the
string tension of the full theory \cite{dfgo96}. Nonetheless, in the opposite
extreme case $SU(N)\simeq SU(N)/Z(N)$ one obtains\ a system with phase
structure quite similar to the previous case. Thus, the action $\mathcal{S}%
_{\mu\nu}\left(  x\right)  $ is split into $\widetilde{\mathcal{S}}_{\mu\nu
}\left(  x\right)  $ with sharp maximum at $\chi=N$ ($\varphi_{\mu\nu}\left(
x\right)  =\varphi_{\mu\nu}^{eff}=0$) and polynomial in $\exp\left\{
-2\hat{\beta}_{\nu\mu}\right\}  $ with supremum at $\chi=0$. Therefore,
maximum $\mathcal{S}_{\mu\nu}\left(  x\right)  $ is located between $\chi=N$
and $\chi=0$ and for any small, but finite difference $\pi/N-\left|
\varphi_{\mu\nu}^{eff}\right|  $ maximum $\mathcal{S}_{\mu\nu}\left(
x\right)  $ steadily moves to $\chi=N$ with increasing $\beta$. Really, it is
enough to have $\chi>0$ to get good grounds for discarding the terms
$\exp\left\{  -2\hat{\beta}_{\nu\mu}\right\}  $ for $\beta\rightarrow\infty$
and in this case we obtain a version of positive plaquette action%

\begin{equation}
-\mathcal{S}_{\mu\nu}\left(  x\right)  =-\widetilde{\mathcal{S}}_{\mu\nu
}\left(  x\right)  +O\left(  e^{-2\hat{\beta}_{\mu\nu}}\right)  =\left|
\hat{\beta}_{\mu\nu}\right|  -\beta+O\left(  e^{-2\hat{\beta}_{\mu\nu}%
}\right)
\end{equation}

Similar models from a more general point of view were intensively studied in
recent years (see e.g. \cite{pos-plaq}). It was shown, that such approximation
didn't change the continuum limit, i.e.\/, the universality class. Moreover,
the Callan-Symansik $\beta$--function of the positive plaquette model shows no
annoying ``dip'' inherent to standard Wilson action \cite{pos-plaq}. Thereby,
as it can be anticipated, for $\beta\rightarrow\infty$ the center subgroup
contributes only exponentially small terms.

Now let's consider a more general case. As it is known, the lattice action of
a given continuum theory is not unique and one could consider an extended
lattice theory, that includes higher representations and belongs to the same
universality class \cite{GGM-94}. The study of the phase diagram of
fundamental-adjoint pure gauge systems revealed a non-trivial and considerably
more complicated phase structure \cite{GL81-B82}. So we consider a more
general case where the plaquette action $\mathcal{S}_{\mu\nu}$ includes an
arbitrary set of irreducible representations $j\equiv\left\{  l_{1}%
,l_{2},...,l_{N-1}\right\}  $\footnote{Integer numbers $l_{n}$ obey
$l_{n+1}<l_{n}$.}
\begin{equation}
-\mathcal{S}_{\mu\nu}\left(  x\right)  =\sum_{j}\beta_{j}\left(  \chi
_{j}\left(  \varphi\right)  -\chi_{j}\left(  0\right)  \right)  ;\quad
\beta_{j}=\beta\eta_{j}\label{ac}%
\end{equation}
with $\eta_{j}=const$ and $\chi_{j}\left(  \varphi\right)  =\mathrm{Tr}%
\left\{  U_{\nu\mu}^{\left(  j\right)  }\left(  x\right)  \right\}  $. Here
plaquette variables $U_{\nu\mu}^{\left(  j\right)  }\left(  x\right)  $ are
expressed in the same way as fundamental ones in $\left(  \ref{p}\right)  $
through the link variables $U_{\nu}^{\left(  j\right)  }\left(  x\right)  $%
\begin{equation}
U_{\nu}^{\left(  j\right)  }\left(  x\right)  =\exp\left\{  i\widehat{\varphi
}_{\nu}^{\left(  j\right)  }\left(  x\right)  \right\}  ;\quad\widehat
{\varphi}_{\nu}^{\left(  j\right)  }\left(  x\right)  \equiv\sum_{m=1}%
^{N^{2}-1}\varphi_{m;\nu}\left(  x\right)  T_{m}^{\left(  j\right)
}\label{Udef}%
\end{equation}
where $SU(N)$ matrices $T_{m}^{\left(  j\right)  }$ are the group generators
in irreducible representations $j$, that obey
\begin{equation}
\mathrm{Tr}\left(  T_{n}^{\left(  j\right)  }T_{m}^{\left(  j\right)
}\right)  =\delta_{nm}\frac{\mathrm{Tr}I^{\left(  j\right)  }}{N^{2}-1}%
C_{2}\left(  j\right) \label{j-r}%
\end{equation}
where $I^{\left(  j\right)  }$ is the unite matrix and $C_{2}\left(  j\right)
$ is the quadratic Casimir operator. For instance, in $SU(3)$ case Casimir
operator is $C_{2}\left(  j\right)  =\left(  l_{1}^{2}+l_{2}^{2}-l_{1}%
l_{2}\right)  /3-1$.

Since the center subgroup contribution is neglected, action $S$\ has single
(up to a gauge transformation) minimum. If we fix the gauge having put
$U_{\mu}^{\left(  j\right)  }\left(  x\right)  =1$ for 'redundant' links, this
minimum will located at $\varphi_{m;\nu}=0$. Since the minimum point is
nondegenerate with $S$\ being infinitely differentiable, $\operatorname{Re}%
S>0$\ and $\operatorname{Im}S=0$, the conditions are fulfilled to make it
possible to apply the Laplace method for the integral $\left(  \ref{parti}%
\right)  $\ evaluation and the result for partition function computation may
be written immediately\textrm{. }Yet we prefer to compute it gradually,
estimating the errors introduced in each step. To begin we expand $\chi
_{j}\left(  \varphi\right)  $ in the minimum point of action $\left(
\ref{ac}\right)  $
\begin{equation}
\chi_{j}\left(  \varphi\right)  \simeq\chi_{j}\left(  0\right)  -\frac
{\chi_{j}\left(  0\right)  C_{2}\left(  j\right)  }{2\left(  N^{2}-1\right)
}\varphi_{\nu\mu}\left(  x\right)  ^{2}+O\left(  \varphi^{4}\right)
;\quad\chi_{j}\left(  0\right)  =\mathrm{Tr}I^{\left(  j\right)  }\label{pj}%
\end{equation}
with%

\begin{equation}
\varphi_{\nu\mu}\left(  x\right)  ^{2}\equiv\sum_{n=1}^{N^{2}-1}\left(
\varphi_{n;\mu}\left(  x\right)  +\varphi_{n;\nu}\left(  x+\mu\right)
-\varphi_{n;\mu}\left(  x+\nu\right)  -\varphi_{n;\nu}\left(  x\right)
\right)  ^{2}%
\end{equation}
and finally get
\begin{equation}
-\mathcal{S}_{\mu\nu}\left(  x\right)  \simeq-\tfrac{\kappa\beta}{2}%
\varphi_{\nu\mu}\left(  x\right)  ^{2}\label{m}%
\end{equation}
with
\begin{equation}
\kappa=\frac{1}{N^{2}-1}\sum_{j}\eta_{j}\chi_{j}\left(  0\right)  C_{2}\left(
j\right) \label{kap}%
\end{equation}
Thus, in the area of asymptotically large $\beta$ the higher representations
contribution leads to plain renormalization of coupling $\beta\rightarrow
\kappa\beta$.

Action $\left(  \ref{m}\right)  $, known as the Manton action \cite{manton},
originally is defined as
\begin{equation}
S_{M}=\beta d^{2}\left(  U_{\mu\nu},I\right)
\end{equation}
where $d\left(  U_{1},U_{2}\right)  $ is the interval between $U_{1}$ and
$U_{2}$ in group space, that for $SU(2)$ may be written as%

\begin{equation}
d\left(  U_{1},U_{2}\right)  =\arccos\tfrac{1}{2}\mathrm{Tr}\left(  U_{1}%
U_{2}^{\dagger}\right)  ,
\end{equation}
and we get%

\begin{equation}
\cos\left[  d\left(  U_{\mu\nu},I\right)  \right]  =\tfrac{1}{2}%
\mathrm{Tr}\left(  U\right)  =\tfrac{1}{2}\chi=\cos\varphi_{\mu\nu}%
\end{equation}
finally coming to $\left(  \ref{m}\right)  $.

In \cite{bcp84} solid grounds are given to suppose that Wilson and Manton
actions belong to the same universality class, so Manton action may be
regarded as a suitable alternative action with correct continuum limit
\cite{MT88}. Furthermore, in \cite{MT88} weighty arguments are presented in
favor of Manton action providing an appreciably faster approach to the
continuum limit than does the Wilson's. Moreover, Manton action violates
asymptotic scaling in the same direction as does the standard one, but is
significantly more weakly \cite{MT88}.

As it was shown in \cite{gk81}, Manton action violates Osterwalder-Schrader
positivity condition \cite{oss} essential for the continual theory. Such
violation, however, appears only in the strong coupling region $\left(
\beta\ll1\right)  $. Indeed, the positivity condition may be written as
\begin{equation}
\int F\left(  U_{1}\right)  \exp\left\{  -\beta d^{2}\left(  U_{1}%
,U_{2}\right)  \right\}  G\left(  U_{2}\right)  d\mu\left(  U_{1}\right)
d\mu\left(  U_{2}\right)  \geq0
\end{equation}
that in a case of $SU(2)$ ($\varphi_{1}=-\varphi_{2}=\phi/2$) is equivalent to
positivity requirement of all coefficients
\begin{equation}
\zeta_{j}^{M}\equiv\int_{-2\pi}^{2\pi}e^{-\beta\phi^{2}/8}\chi_{j}d\mu
=\frac{e^{-2j^{2}/\beta}-e^{-2\left(  j+1\right)  ^{2}/\beta}}{\sqrt{2\pi
\beta}}\left(  1+O\left(  e^{-\pi^{2}\beta/2}\right)  \right) \label{kM}%
\end{equation}
and in case of asymptotically large $\beta$ Osterwalder-Schrader positivity
condition is fulfilled.

The expression $\left(  \ref{kM}\right)  $ allows to roughly estimate the
error of considered approximation. If we compare $\zeta_{j}^{M}$ coefficients
with those computed for Wilson action
\begin{equation}
\zeta_{j}^{W}\equiv\int e^{\beta\left(  \cos\frac{\phi}{2}-1\right)  }\chi
_{j}d\mu=\left(  2j+1\right)  e^{-\beta}I_{2j+1}\left(  \beta\right)  /\beta
\end{equation}
one can easily show that $\zeta_{j}^{W}/\zeta_{j}^{M}=1+O\left(
1/\beta\right)  $.

Now the measure, that in general is defined as%

\begin{equation}
d\mu=\sqrt{\det_{\left(  kn\right)  }\mathrm{Tr}\left\{  \frac{\partial
U}{\partial\varphi_{k}}\frac{\partial U^{\dagger}}{\partial\varphi_{n}%
}\right\}  }\prod_{m=1}^{N^{2}-1}d\varphi_{m}%
\end{equation}
may be computed. The integrand in $\left(  \ref{parti}\right)  $ has a sharp
maximum at $\varphi_{\nu\mu}\left(  x\right)  =0$, meaning that (up to gauge
transformation) it has acute maximum at $\varphi_{n;\nu}\left(  x\right)  =0$.
Hence computing the measure with the same accuracy as in $\left(
\ref{pj}\right)  $, one may get
\begin{equation}
d\mu\simeq C\exp\left\{  -\sum_{n=1}^{N^{2}-1}\varphi_{n}\varphi_{n}/\left(
N^{2}-1\right)  \right\}  \prod_{n=1}^{N^{2}-1}d\varphi_{n};\qquad N=2,3.
\end{equation}
Therefore, we may finally write for the partition function
\begin{equation}
Z\simeq\int\exp\left\{  -\beta\left(  1+\tfrac{1}{N^{2}-1}\right)  \sum
_{\nu\mu,x}\varphi_{\nu\mu}\left(  x\right)  ^{2}/2\right\}  \prod_{\nu
,x}\prod_{n=1}^{N^{2}-1}d\varphi_{n,\nu}\left(  x\right) \label{ZM}%
\end{equation}
since the measure contribution had been very important in the region
$\beta\sim N$\ , in the area $\beta\gg1$ it\ introduces negligible (of order
$1/\beta$) correction.

As long as the integration area is compact $\left|  \varphi\right|  {}%
<\varphi{}_{\text{supr}}$, theory remains non-trivial and, at least, on finite
lattice shows critical behavior at $g\sim1$ \cite{MT88}. Since the action
$\left(  \ref{m}\right)  $ yields a real positive-definite quadratic form, the
error introduced by the extension of area integration by $\varphi_{n;\mu
}\left(  x\right)  $ to infinity is of the order $\exp\left\{  -\beta
\varphi^{2}{}_{\text{supr}}/2\right\}  $. As a matter of fact, such extension
is doubtful at finite $\beta$, but positively harmless for $\beta\gg1$. So at
asymptotically large $\beta,$ integration in $\left(  \ref{ZM}\right)  $ may
be done trivially by substitution $\varphi_{n;\mu}\left(  x\right)
\rightarrow\varphi_{n;\mu}\left(  x\right)  /\sqrt{\beta\kappa}$ and we
immediately get
\begin{equation}
\digamma=-\frac{1}{N_{\tau}N_{\sigma}^{3}}\ln Z\left(  \beta\kappa\right)
\simeq-C\ln\frac{\beta_{1}}{\beta}\label{fr}%
\end{equation}
where $\beta_{1}$ is defined by
\begin{equation}
C\ln\beta_{1}=\frac{1}{N_{\tau}N_{\sigma}^{3}}\ln Z\left(  1\right)
\end{equation}
Coefficient $C$ includes a factor to account for the fact that the number of
integration variables is by approximately a quarter less than that of the
links, because in order to fix the gauge we must freeze 'redundant' link
variables in$\left(  \ref{m}\right)  $, i.e. put $\widehat{\varphi}_{\sigma
}^{\left(  j\right)  }\left(  x\right)  =0$ for corresponding link variables.

\section{Fermion contribution}

In perturbation theory the fermion part of action doesn't play a leading role
in computing of the Callan-Symanzik $\beta$-function $~\beta_{CS}(g_{0})$,
nonetheless, its contribution is discernible, especially in three-loop
calculation \cite{cfpv}. Since the growing importance of the fermion
contribution in nonperturbative calculations can't be excluded for
$\beta\rightarrow\infty$, such input should be, at least, roughly estimated.

Unfortunately, we can hardly attack the problem in its full, hence the
approximations which hopefully capture some of the essential features of the
physics may be considered. We attempt to study the fermionic action (see e.g.
\cite{HK})
\begin{equation}
-S_{F}\equiv-\sum_{\mathbf{x}}\mathcal{S}_{F}\left(  \mathbf{x}%
\right)  =n_{f}a^{3}\sum_{x,x^{\prime}}\left(  \overline{\psi}_{x^{\prime}%
}D_{x^{\prime}x}^{0}\psi_{x}+\widetilde{\xi}^{-1}\!\!\!\ \overline{\psi
}_{x^{\prime}}\sum_{n=1}^{3}D_{x^{\prime}x}^{n}\psi_{x}\right) \label{sf}%
\end{equation}
with
\begin{equation}
D_{x^{\prime}x}^{\nu}=\tfrac{1-\gamma_{\nu}}{2}U_{\nu}\left(  x\right)
\delta_{x,x^{\prime}-\nu}+\tfrac{1+\gamma_{\nu}}{2}U_{\nu}^{\dagger}\left(
x^{\prime}\right)  \delta_{x,x^{\prime}+\nu}-\left(  1+\delta_{\nu}%
^{0}ma_{\tau}\right)  \delta_{x^{\prime},x},\label{D}%
\end{equation}
on an extremely anisotropic lattice ($\widetilde{\xi}\gg1$) in the
approximation where the terms proportional to $\widetilde{\xi}^{-1}$ are
discarded \cite{p}. Here $\gamma_{\nu}$ are Dirac matrices, $n_{f}$\ is the
number of flavors and $\widetilde{\xi}=$ $\widetilde{\xi}\left(  g,\xi\right)
$ is the '\textit{bare}' anisotropy parameter. The dependance of
$\widetilde{\xi}$ on coupling $g$ and '\textit{renormalized' }anisotropy
parameter $\xi=a/a_{\tau}\ $is defined by the condition of independence of
physical values on spatial $a$ and temporal $a_{\tau}$ lattice spacings.

Fermion action $S\!\!\!\!_{F}$ doesn't depend on $g$ explicitly, nonetheless,
such dependence may be induced through $\widetilde{\xi}$, because on the
anisotropic lattice it enters in Yang-Mills part of action as well (temporal
and spatial part is directly and inversely proportional to $\widetilde{\xi}$,
respectively). However, there are some reasons to believe that such dependence
quickly disappears with $g\rightarrow0$. Indeed, recent analysis \cite{eks99}
shows that, at least, for $1.5\leq\xi\leq6$\ function $\widetilde{\xi}\left(
g,\xi\right)  $\ is linear in $\xi$ (with the natural condition of
$\widetilde{\xi}\left(  g,1\right)  =1$)
\begin{equation}
\widetilde{\xi}\left(  g,\xi\right)  \simeq\xi+\,\allowbreak q\left(
g\right)  \left(  1-\xi\right)  .\label{ksi}%
\end{equation}
The comparison of $\left(  \ref{ksi}\right)  $ with the data in \cite{k98}
allows to conclude that $\,\allowbreak q\left(  g\right)  $ steadily decreases
with $g\rightarrow0$ and may be fitted 'on eye' as%

\begin{equation}
q\left(  g\right)  \simeq.0\,\allowbreak2\,-.0\,\allowbreak2\allowbreak
g^{2}+.\,\allowbreak3\allowbreak g^{4}+O\left(  g^{6}\right)  ;
\end{equation}
thereby the dependance $\widetilde{\xi}$ on $g$ becomes inessential for small
$g$. It allows us to assume that such dependance may be ignored for
$g\rightarrow0$ even in the area of large $\xi$, all the more that
$S\!\!\!\!_{F}$ doesn't depend on $g$ explicitly.

We fix diagonal static Hamiltonian gauge
\begin{equation}
U_{0}\left(  \mathbf{x},t\right)  _{\nu\mu}=\left\{
\begin{array}
[c]{ccc}%
\delta_{\nu\mu} & for & t\neq0\\
\delta_{\nu\mu}\exp\left\{  i\phi_{\nu}\left(  \mathbf{x}\right)  \right\}  &
for & t=0
\end{array}
\right.  ;
\end{equation}
where $\phi_{\nu}\left(  \mathbf{x}\right)  $ e.g. for $N=3$ is given by
\begin{equation}
\phi_{1,2}\left(  \mathbf{x}\right)  =\pm\varphi_{3,0}\left(  \mathbf{x}%
,0\right)  +\varphi_{8,0}\left(  \mathbf{x},0\right)  /\sqrt{3};\qquad\phi
_{3}\left(  \mathbf{x}\right)  =-\phi_{1}\left(  \mathbf{x}\right)  -\phi
_{2}\left(  \mathbf{x}\right)  =-2\varphi_{8}/\sqrt{3}%
\end{equation}
and after integration of partition function over fermion fields $\psi_{x}$ we
get \cite{p}
\begin{equation}
-\mathcal{S}_{F}\left(  \mathbf{x}\right)  =\sum_{\alpha=1}^{3}\ln\left(
\cos\phi_{\alpha}+\cosh\tfrac{m}{T}\right)  +O\left(  \xi^{-2}\right)
\end{equation}
that gives (up to the additive constant)
\begin{equation}
-\mathcal{S}_{F}\left(  \mathbf{x}\right)  =-\frac{\varphi_{3,0}^{2}\left(
\mathbf{x},0\right)  +\varphi_{8,0}^{2}\left(  \mathbf{x},0\right)  }%
{1+\cosh\frac{m}{T}}+O\left(  \varphi^{4}\right)  =-\frac{\varphi_{0}%
^{2}\left(  \mathbf{x},0\right)  }{1+\cosh\frac{m}{T}}+O\left(  \varphi
^{4}\right)
\end{equation}
So, we see that the fermion part includes only temporal link variables and
yields a real positive-definite quadratic form as well as $\left(
\ref{m}\right)  $.\ However, unlike $\left(  \ref{m}\right)  $, $S_{F}$\ does
not depend on $\beta$\ explicitly and \ plays a minor part for $\beta
\rightarrow\infty$.

$\allowbreak\allowbreak$We would stress that fermion contribution into the
action depends on $N_{\tau}$\ and $a$\ only in $N_{\tau}a=1/T$\ combination
and one may hope that this property will survive in exact solution on
isotropic lattices. In this case the fermion part, that doesn't depend on
$\beta$ explicitly, becomes negligible in comparison with $\left(
\ref{m}\right)  $ for $\beta\rightarrow\infty$.

\section{Conclusions}

We must conclude that in the area of extremely large $\beta$ the theory
becomes trivial and partition function doesn't show any sign of critical
behavior. As it follows from $\left(  \ref{fr}\right)  $, the condition of
finiteness of free energy density%

\begin{equation}
F=a^{-4}T\digamma=-\tfrac{T}{aV}\ln Z\propto a^{-4}T\ln\frac{\beta}{\beta_{1}%
};\quad V=\left(  aN_{\sigma}\right)  ^{3}%
\end{equation}
in continuum limit leads to
\begin{equation}
\ln\frac{\beta}{\beta_{1}}\propto a^{4}%
\end{equation}
which strongly contradicts to $\left(  \ref{bet}\right)  $, but doesn't
disagree in substance with fixed point model predictions.

To complete the picture let us consider partition function behavior for
$\beta\rightarrow\infty$ in $SU\left(  N\right)  \simeq Z\left(  N\right)  $
approximation. Since on the dual lattice only surfaces of small area survive,
the first non-trivial contribution comes from a six-plaquette surface (a cube)
so, making allowance for $\left(  \ref{zN}\right)  $, $\left(  \ref{dc}%
\right)  $ and $\left(  \ref{kc-F}\right)  $, we may write for the partition
function
\begin{equation}
\ln Z\simeq-\Xi=N_{\sigma}^{3}N_{\tau}\exp\left\{  -12\beta\right\}
\label{Z--Z(N)}%
\end{equation}
Although free energy density also behaves \textit{trivially}
\begin{equation}
F=-V^{-1}T\ln Z\propto a^{-4}T\exp\left\{  -12\beta\right\}  ,
\end{equation}
this, however, unexpectedly leads to passable agreement with $\left(
\ref{bet}\right)  $ in continuum limit
\begin{equation}
g^{-2}\simeq0.1\ln\tfrac{1}{\Lambda_{0}a};\quad\Lambda_{0}=const.
\end{equation}

On the other hand, the presence of the critical point $\beta_{c}<\infty$ at
finite $N_{\tau}$ is well-established in MC experiment. So, we have to
conclude, that with $N_{\tau}\rightarrow\infty$ the $\beta_{c}$ critical point
either expires, or gradually approaches some \textit{finite} value.

Equation $\left(  \ref{bet}\right)  $ is obtained in continuum field theory,
that differs from LGT, at least, in two essential points: it is noncompact and
there is no local distinction between pure $SU(N)$\ and $SU(N)/Z(N)$
\cite{kt98}. The arguments presented above allow us to think, that the
last-mentioned difference quickly enough disappears in the region of
asymptotically large $\beta$. Compact and noncompact formulations of the
theory differ by the term of $O\left(  \exp\left\{  -\beta\right\}  \right)  $ order.

In addition, as it can be seen from $\left(  \ref{kap}\right)  $, coefficient
$\kappa$ may be incorporated in $\beta$ by simple constant renormalization, so
for $\beta\rightarrow\infty$ the action $\left(  \ref{m}\right)  $ is
insensible to any set of irreducible representations entered in original
action $\left(  \ref{ac}\right)  $.

Unfortunately, as it follows from $\left(  \ref{m}\right)  $, with increasing
$\beta$ the action looses its sensitivity to nonabelian properties of the
gauge group, that with $d\mu\simeq d\varphi$ approximation reduces the gauge
group $SU\left(  N\right)  $ to $U\left(  1\right)  ^{N^{2}-1}$. One can
hardly expect such 'simplification' in field theory even for $\beta
\rightarrow\infty$.

Let us finally list the approximations considered in the area $\beta
\rightarrow\infty$

\begin{enumerate}
\item  Center group contribution is neglected as exponentially small.

\item  Yang-Mills part of the action is expanded into power series at the
maximum and only quadratic terms are preserved.

\item  Measure introduces into effective action the corrections of order
$1/\beta$, that may be neglected for $\beta\rightarrow\infty$.

\item  Fermion term is neglected. It may be partially justified by the fact,
that on an extremely anisotropic lattice such term doesn't depend on $\beta$
explicitly and therefore is actually negligible for $\beta\rightarrow\infty$.

\item  Integration area is extended to infinity, that introduces an error of
$\exp\left(  -\beta\right)  $ order.
\end{enumerate}

\end{document}